\documentclass[11pt,letterpaper]{article}

\usepackage{times,mathptmx}
\usepackage[small,it]{caption} 

\usepackage{fullpage}
\usepackage{multirow}
\usepackage{amsmath}
\usepackage{amssymb}
\usepackage{amsthm}
\usepackage{hyperref}
\usepackage{subfigure}

\usepackage{tikz}
\usetikzlibrary{arrows,decorations,decorations.shapes,backgrounds,shapes}

\newtheorem{theorem}{Theorem}[section]
\newtheorem{lemma}[theorem]{Lemma}

\newtheorem{conjecture}[theorem]{Conjecture}

\theoremstyle{definition}

\newcommand{\E}[1]{\mathbb{E}\left[\;#1\;\right]}
\renewcommand{\Pr}[1]{\text{Pr}\left[\;#1\;\right]}
\renewcommand{\vec}{\mathbf}
\newcommand{\opt}{\text{OPT}}
\newcommand{\mech}{\text{MECH}}
\newcommand{\Gcal}{\mathcal{G}}
\newcommand{\Scal}{\mathcal{S}}
\renewcommand{\deg}{\text{deg}}
\newcommand{\bigO}{\mathcal{O}}

\begin{document}

\title{Sum of Us: Strategyproof Selection from the Selectors}
\author{%
Noga Alon\thanks{%
	Microsoft Israel R\&D Center,
	13 Shenkar Street, Herzeliya 46725, Israel,
	and Schools of Mathematics and Computer Science,  
	Tel Aviv University, Tel Aviv, 69978, Israel,
	Email: \texttt{nogaa@tau.ac.il}.
	Research supported in part by a USA Israeli BSF grant, by a grant from
the Israel Science Foundation, by an ERC advanced grant and by the Hermann Minkowski Minerva Center for Geometry at Tel Aviv University.}
\and
Felix Fischer\thanks{%
	Institut f{\"u}r Informatik,
  Ludwig-Maximilians-Universit{\"a}t M{\"u}nchen,
  80538 M{\"u}nchen, Germany, 
  email: \texttt{fischerf@tcs.ifi.lmu.de}. 
  The author is supported by the Deutsche Forschungsgemeinschaft under grant BR~2312/3-2. Part of the work was done during a visit to The Hebrew University of Jerusalem, which was supported by a Minerva Short-Term Research Grant.}
\and 
Ariel D.~Procaccia\thanks{%
	School of Engineering and Applied Sciences,
	Harvard University, Cambridge, MA 02138,  
	Email: \texttt{arielpro@seas.harvard.edu}.
	This work was done while the author was at Microsoft Israel R\&D Center.}
\and
Moshe Tennenholtz \thanks{%
	Microsoft Israel R\&D Center,
	13 Shenkar Street, Herzeliya 46725, Israel,
	and Technion, IIT, 
	Haifa 32000, Israel.  
	Email: \texttt{moshet@microsoft.com}}}
\date{}
\maketitle

\begin{abstract}
	We consider directed graphs over a set of $n$ agents, where an edge $(i,j)$ is taken to mean that agent $i$ supports or trusts agent $j$. Given such a graph and an integer $k\leq n$, we wish to select a subset of $k$ agents that maximizes the sum of indegrees, i.e., a subset of $k$ most popular or most trusted agents. At the same time we assume that each individual agent is only interested in being selected, and may misreport its outgoing edges to this end. This problem formulation captures realistic scenarios where agents choose among themselves, which can be found in the context of Internet search, social networks like Twitter, or reputation systems like Epinions. 
	
	Our goal is to design mechanisms without payments that map each graph to a $k$-subset of agents to be selected and satisfy the following two constraints: \emph{strategyproofness}, i.e., agents cannot benefit from misreporting their outgoing edges, and \emph{approximate optimality}, i.e., the sum of indegrees of the selected subset of agents is always close to optimal. Our first main result is a surprising impossibility: for $k\in\{1,\ldots,n-1\}$, no deterministic strategyproof mechanism can provide a finite approximation ratio. Our second main result is a randomized strategyproof mechanism with an approximation ratio that is bounded from above by four for any value of $k$, and approaches one as $k$ grows. 
\end{abstract}

\setcounter{page}{0} \thispagestyle{empty} 
\newpage

\section{Introduction}
\label{sec:intro}

One of the most well-studied settings in social choice theory concerns a set of \emph{agents} (also known as \emph{voters} or \emph{individuals}) and a set of \emph{alternatives} (also known as \emph{candidates}). The agents express their preferences over the alternatives, and these are mapped by some function to a winning alternative or set of winning alternatives. In one prominent variation, each agent must select a subset of alternatives that it approves; this setting is known as \emph{Approval voting}~\cite{BF07}. 

We consider the special case of Approval voting when the set of agents and the set of alternatives coincide. Specifically, in our model there is an underlying directed graph, with the agents as vertices. An edge from agent $i$ to agent $j$ implies that agent $i$ approves, votes for, trusts, or supports agent $j$. Our goal is to select a subset of $k$ ``best'' agents, based on the given graph; we elaborate on what we mean by ``best'' momentarily.

Our assumption that agents and alternatives coincide enables us to restrict the agents' preferences. Indeed, we assume that each agent is only interested in whether it is among those selected, that is, it receives utility one if selected and zero otherwise. This assumption reflects, in the limit, a situation where each agent gives very small weight to the overall composition of the selected subset, and very high weight to the question of its own selection.\footnote{See Section~\ref{sec:disc} for further discussion of this utility model.}

As a first motivating example, consider an Internet search setting. The web sites are the agents, while hyperlinks are represented by edges. Given this graph, a search engine must return a set of the, say, ten top web sites. Put another way, the top web sites are selected based on the votes cast by other web sites in the form of hyperlinks. Each specific web site or, more accurately, its webmaster is naturally concerned only with appearing at the top of the search results, and to this end may add or remove hyperlinks at will.

A (deterministic) \emph{$k$-selection mechanism} is a function that maps a given graph on the set of agents to a $k$-subset of selected agents. We also consider \emph{randomized} $k$-selection mechanisms, which randomly select a subset. 

The outgoing edges in the underlying graph $G$ are private information of the respective agent. Fixing a mechanism $f$, the agents play the following game. Each of them reports to the mechanism a set of outgoing edges, which might differ from the true set. The reported edges induce a graph $G'$, and the mechanism selects the subset $f(G')$. We say that a mechanism is \emph{strategyproof (SP)} if an agent cannot benefit from misreporting its outgoing edges, that is, cannot increase its chances of being selected, even if it has complete information about the rest of the graph. We further say that a mechanism is \emph{group strategyproof (GSP)} if even a coalition of agents cannot all gain from misreporting their outgoing edges. 

We now explain what we mean by selecting the ``best'' agents. In this paper, we measure the quality of a set of agents by their total number of incoming edges, i.e., the sum of their indegrees. The goal of the mechanism designer is to optimize this target function. Note that this goal is in a sense orthogonal to the agent's interests, which may make the design of good SP mechanisms difficult. 

A second motivating example can be found in the context of social networks. While some social networks, like Facebook (\url{http://facebook.com}), correspond to undirected graphs, there are many examples with unilateral connections. Each user of the reputation system Epinions (\url{http://epinions.com}) has a ``Web of Trust'', that is, the user unilaterally chooses which other users to trust. Another prominent example is the social network Twitter (\url{http://twitter.com}), which of late has become wildly popular; a Twitter user may choose which other users to ``follow.''

In ``directed'' social networks, choosing a $k$-subset with maximum overall indegree simply means selecting the $k$ most popular or most trusted users. Applications include setting up a committee, recommending a trusted group of vendors, targeting a group for an advertising campaign, or simply holding a popularity contest. The last point may seem pure fantasy, but, indeed, celebrity users of Twitter have recently held a race to the milestone of one million followers; the dubious honor ultimately went to actor Ashton Kutcher. Clearly Mr.~Kutcher could increase the chance of being selected by not following any other users, that is, reporting an empty set of outgoing edges.

Since a mechanism that selects an optimal subset (in terms of total indegree) is clearly not SP, we will resort to approximate optimality. More precisely, we seek SP mechanisms that give a good approximation, in the usual sense, to the total indegree. Crucially, approximation is \emph{not} employed in this context to circumvent computational complexity (as the problem of selecting an optimal subset is obviously tractable), but in order to sufficiently broaden the space of acceptable mechanisms to include SP ones. 

\bigskip
\noindent\textbf{Context and related work.}
The work in this paper falls squarely into the realm of \emph{approximate mechanism design without money}, an agenda recently introduced by some of us (Procaccia and Tennenholtz~\cite{PT09}), building on earlier work (for example by Dekel et al.~\cite{DFP08}). This agenda advocates the design of SP approximation mechanisms \emph{without payments} for structured, and preferably computationally tractable, optimization problems. Indeed, while almost all the work in the field of \emph{algorithmic mechanism design}~\cite{NR01} considers mechanisms that are allowed to transfer payments to and from the agents, money is usually unavailable in Internet domains like the ones discussed above (social networks, search engines) due to security and accountability issues (see, e.g., the book chapter by Schummer and Vohra~\cite{SV07}). Our notion of a mechanism, sometimes referred to as a \emph{social choice rule} in the social choice literature, therefore precludes payments by definition. Note that Procaccia and Tennenholtz~\cite{PT09}, and also Alon et al.~\cite{AFPT09}, deal with a completely different domain, namely facility location.

For $k=1$, that is, if one agent must be selected, the game we deal with is a special case of so-called \emph{selection games}~\cite{AT08}, where the possible strategies are the outgoing edges. More generally, this setting is related to work in distributed computing on \emph{leader election} (see, e.g.,~\cite{AN93,CL95,Fei99,Ant06}). This line of work does not deal with self-interested agents.  Instead, there is a certain number of malicious agents trying to manipulate the selection process, and the goal is to guarantee the selection of a non-malicious agent, at least with a certain probability.

Finally, this paper is related to work on manipulation of reputation systems, which are often modeled as weighted directed graphs; a reputation function maps a given graph to reputation values for the agents (see, e.g.,~\cite{CF05,FRS07}). Although our positive results can be extended to weighted graphs, when the target function is the sum of weights on incoming edges, this would hardly be a reasonable target function. Indeed, in this context the absence of a specific incoming edge (which indicates lack of knowledge) is preferable to an edge with low weight (which indicates distrust); see Section~\ref{sec:disc} for further discussion.

\bigskip
\noindent\textbf{Our results and techniques.} 
We give rather tight upper and lower bounds on the approximation ratio achievable by $k$-selection mechanisms in the setting described above; the properties of the mechanisms fall along two orthogonal dimensions: deterministic vs.\@ randomized, and SP vs.\@ GSP. A summary of our results is given in Table~\ref{tab:results}. 
\begin{table}[t]
	\centering
	\renewcommand{\arraystretch}{1.2}
	\setlength{\tabcolsep}{2pt}
	\begin{tabular}{|c|c|c|c|}
	\cline{3-4}
	\multicolumn{2}{c|}{} & \textbf{Deterministic} & \textbf{Randomized} \\
	\hline
	\multirow{2}{*}{\textbf{SP}} & \textbf{Upper bound} & n/a & $\min\{4,1+\bigO(1/k^{1/3})\}$ \\
	\cline{2-4}
	& \textbf{Lower bound} & $\infty$ & $1+\Omega(1/k^2)$ \\ 
	\hline\hline
	\multirow{2}{*}{\textbf{GSP}} & \textbf{Upper bound} & n/a & $\frac{n}{k}$ \\
	\cline{2-4}
	& \textbf{Lower bound} & $\infty$ & $\frac{n-1}{k}$ \\ 
	\hline
	\end{tabular}
\caption{Summary of our results for $k$-selection mechanisms, where $n$ is the number of agents. SP stands for strategyproof, GSP for group strategyproof.}
\label{tab:results}
\end{table}

Our contribution begins in Section~\ref{sec:det} with a study of deterministic $k$-selection mechanisms. It is quite easy to see that no deterministic SP $1$-selection mechanism can yield a finite approximation ratio. Intuitively, this should not be true for large values of $k$. Indeed, in order to have a finite approximation ratio, a mechanism should very simply select a subset of agents with at least one incoming edge, if there is such a set. In the extreme case when $k=n-1$, we must select all the agents save one, and the question is whether there exists an SP mechanism that never eliminates the unique agent with positive indegree. Our first result gives a surprising negative answer to this question, and in fact holds for every value of $k$. 

\medskip
\noindent\textbf{Theorem~\ref{thm:det_lb}.} \emph{Let $N=\{1,\ldots,n\}$, $n\geq 2$, and $k\in\{1,\ldots,n-1\}$. Then there is no deterministic SP $k$-selection mechanism that gives a finite approximation ratio.}
\medskip 

The proof of the theorem is compact but rather tricky. It involves two main arguments. We first restrict our attention to a subset of the graphs, namely to stars with all edges directed at a specific agent. An SP mechanism over such graphs can be represented using a function over the boolean $(n-1)$-cube, which must satisfy certain constraints. We then use a parity argument to show that the constraints lead to a contradiction.

In Section~\ref{sec:rand} we turn to randomized $k$-selection mechanisms. We design a randomized mechanism, \emph{Random $m$-Partition ($m$-RP)}, parameterized by $m$, that works by randomly partitioning the set of agents into $m$ subsets, and then selecting the (roughly) $k/m$ agents with largest indegree from each subset, when only the incoming edges from the other subsets are taken into account. This rather simple technique is reminiscent of work on \emph{random sampling} in the context of auctions for digital goods~\cite{FGHK02,GHKS+06,FFHK05} and combinatorial auctions~\cite{DNS06}, although our problem is fundamentally different. We have the following theorem.

\medskip
\noindent\textbf{Theorem~\ref{thm:2group}.} \emph{Let $N=\{1,\ldots,n\}$, $k\in\{1,\ldots,n-1\}$. For every value of $m$, $m$-RP is SP. Furthermore,}
\begin{enumerate}
\item \emph{$2$-RP has an approximation ratio of four, and}
\item \emph{$\left(\left\lceil k^{1/3}\right\rceil\right)$-RP has an approximation ratio of $1+\bigO(1/k^{1/3})$.}
\end{enumerate} 

For a given number $k$ of agents to be selected, we can in fact choose the best value of $m$ when applying $m$-RP. Thus, there exists a mechanism that always yields an approximation ratio of at most four, and furthermore provides a ratio that approaches one as $k$ grows. In addition, we prove a lower bound of $1+\Omega(1/k^2)$ on the approximation ratio that can be achieved by any randomized SP $k$-selection mechanism; in particular, the lower bound is two for $k=1$.

As our final result, we obtain a lower bound of $(n-1)/k$ for randomized GSP $k$-selection mechanisms. This result implies that when asking for group strategyproofness one essentially cannot do better than simply selecting $k$ agents at random, which is obviously GSP and gives an approximation ratio of $n/k$.

\section{The Model}
\label{sec:prem}

Let $N=\{1,\ldots,n\}$ be a set of \emph{agents}. For each $k=1,\ldots,n$, let $\Scal_k=\Scal_k(n)$ be the collection of $k$-subsets of $N$, i.e., $\Scal_k = \{S\subseteq N:\ |S|=k\}$. We consider directed graphs $G=(N,E)$, that is, graphs with $N$ as the set of vertices, and write $\Gcal=\Gcal(N)$ for the set of such graphs. 

A \emph{deterministic $k$-selection mechanism} is a function $f:\Gcal\rightarrow \Scal_k$ that selects a subset of agents for each graph.  When the subset $S\subseteq N$ is selected, agent $i\in N$ obtains utility $u_i(S)=1$ if $i\in S$ and $u_i(S)=0$ otherwise, i.e., agents only care about whether they are selected or not. We further discuss this utility model in Section~\ref{sec:disc}. 

A \emph{randomized $k$-selection mechanism} is a function $f:\Gcal\rightarrow \Delta(\Scal_k)$, where $\Delta(\Scal_k)$ is the set of probability distributions over $\Scal_k$. Given a distribution $\mu\in \Delta(\Scal_k)$, the utility of agent $i\in N$ is
$$
u_i(\mu) = \mathbb{E}_{S\sim \mu} [u_i(S)] = \text{Pr}_{S\sim \mu}[i\in S].
$$
Deterministic mechanisms can be seen as a special case of a randomized ones, always selecting a set of agents with probability one. 

We say that a $k$-selection mechanism is \emph{strategyproof (SP)} if an agent cannot benefit from misreporting its edges. Formally, strategyproofness requires that for every $i\in N$ and every pair of graphs $G,G'\in \Gcal$ that differ only in the outgoing edges of agent $i$, it holds that $u_i(G)=u_i(G')$.\footnote{By symmetry, this is equivalent to writing the last equality as an inequality.} This means that the probability of agent $i\in N$ being selected has to be independent of the outgoing edges reported by $i$. A discussion of this definition in the context of randomized mechanisms can be found in Section~\ref{sec:disc}.

A $k$-selection mechanism is \emph{group strategyproof (GSP)} if there is no coalition of agents that can all gain from jointly misreporting their outgoing edges.  Formally, group strategyproofness requires that for every $S\subseteq N$ and every pair of graphs $G,G'\in\Gcal$ that differ only in the outgoing edges of the agents in $S$, there exists $i\in S$ such that $u_i(G)\leq u_i(G')$. An alternative, stronger definition requires that some agent \emph{strictly} lose as a result of the deviation. Crucially, our result with respect to group strategyproofness is an impossibility, hence using the weaker definition only strengthens the result.

Given a graph $G$, let $\deg(i) = \deg(i,G)$ be the indegree of agent $i$ in $G$, i.e., the number of its incoming edges. We seek mechanisms that are SP or GSP, and in addition approximate the optimization target $\sum_{i\in S} \deg(i)$, that is, we wish to maximize the sum of indegrees of the selected agents. Formally, we say that a $k$-selection mechanism $f$ has an approximation ratio of $\alpha$ if for every graph $G$, 
$$
\frac{\max_{S\in\Scal_k}\sum_{i\in S} \deg(i)}{\mathbb{E}_{S\sim f(G)}[\sum_{i\in S}\deg(i)]}\leq \alpha.
$$

\section{Deterministic Mechanisms}
\label{sec:det}

In this section we study deterministic $k$-selection mechanisms. Before stating our impossibility result, we discuss some special cases. 

Clearly, only one mechanism exists for $k=n$, that is, when all the agents must be selected, and this mechanism is optimal.  More interestingly, it is easy to see that one cannot obtain a finite approximation ratio via a deterministic SP mechanism when $k=1$.  Indeed, let $n\geq 2$, let $f$ be an SP deterministic mechanism, and consider a graph $G=(N,E)$ with $E=\{(1,2), (2,1)\}$, i.e., the only two edges are from agent $1$ to agent $2$ and vice versa. Without loss of generality we may assume that $f(G)=\{1\}$. Now, assume that agent $2$ removes its outgoing edge; formally, we now consider the graph $G'=(N,E')$ with $E'=\{(1,2)\}$. By strategyproofness, $f(G')=\{1\}$, but now agent $2$ is the only agent with positive degree, hence the approximation ratio of $f$ is infinite. 

Note that in order to have a finite approximation ratio, our mechanism must satisfy the following property, which is also sufficient: if there is an edge in the graph, the mechanism must select a subset of agents with at least one incoming edge. The argument above shows that this property cannot be satisfied by any SP mechanism when $k=1$, but intuitively it should be easy to satisfy when $k$ is very large. 

Consider, for example, the case where $k=n-1$, that is, the mechanism must select all the agents save one. Can we design an SP mechanism with the extremely basic property that if there is only one agent with incoming edges, that agent would not be the only one \emph{not} to be selected? 

In the following theorem, we give a surprising negative answer to this question, even when we restrict our attention to graphs where each agent has at most one outgoing edge. Amusingly, a connection to the popular TV game show ``Survivor'' can be made. Consider a slight variation where each tribe member can vote for one other trusted member, but is also allowed not to cast a vote. One member must be eliminated at the tribal council, based on the votes. Since each member's first priority is not to be eliminated (i.e., to be selected), strategyproofness in our $0$--$1$ utility model is in fact a necessary condition for strategyproofness in suitable, more refined utility models. The theorem then implies that a mechanism for choosing the eliminated member cannot be SP (even under $0$--$1$ utilities) if it has the property that a member who is the only one that received votes cannot be eliminated. Put another way, lies are inherent in the game!

More generally, we show that for \emph{any} value of $k$, strategyproofness and finite approximation ratio are mutually exclusive. The proof is concise but nontrivial. 
\begin{theorem}
\label{thm:det_lb}
Let $N=\{1,\ldots,n\}$, $n\geq 2$, and $k\in\{1,\ldots,n-1\}$. There is no deterministic SP $k$-selection mechanism that gives a finite approximation ratio. 
\end{theorem}

\begin{proof}
Assume for contradiction that $f:\Gcal\rightarrow \Scal_k$ is a deterministic SP $k$-selection mechanism that gives a finite approximation ratio. Furthermore, let $G^*=(N,\emptyset)$ be the empty graph. Since $k<n$, there exists $i\in N$ such that $i\notin f(G^*)$; without loss of generality, $n\notin f(G^*)$.

We will restrict our attention to stars whose center is agent $n$, that is, graphs where the only edges are of the form $(i,n)$ for an agent $i\in N\setminus\{n\}$. We can represent such a graph by a binary vector $\vec{x}=(x_1,\ldots,x_{n-1})$, where $x_i=1$ if and only if the edge $(i,1)$ is in the graph; see Figure~\ref{fig:star} for an illustration. In other words, we restrict the domain of $f$ to $\{0,1\}^{n-1}$.

We claim that $n\in f(\vec{x})$ for all $\vec{x}\in \{0,1\}^{n-1}\setminus\{\vec{0}\}$. Indeed, in every such graph agent $n$ is the only agent with incoming edges. Hence, any subset that does not include agent $n$ has zero incoming edges, and therefore does not give a finite approximation ratio (as a subset that does include agent $n$ has at least one incoming edge). 

To summarize, $f$ satisfies the following three constraints: 
\begin{enumerate}
\item $n\notin f(\vec{0})$.
\item For all $\vec{x}\in \{0,1\}^{n-1}\setminus\{\vec{0}\}$, $n\in f(\vec{x})$.
\item Strategyproofness: for all $i\in N\setminus\{n\}$ and $\vec{x}\in \{0,1\}^{n-1}$, $i\in f(\vec{x})$ if and only if $i\in f(\vec{x}+e_i)$, where $e_i$ is the $i$th unit vector and addition is modulo $2$.
\end{enumerate}

Next, we claim that $|\{\vec{x}\in\{0,1\}^{n-1}:\ i\in f(\vec{x})\}|$ is even for all $i\in N\setminus\{n\}$. This follows directly from the third constraint, strategyproofness: we can simply partition the set
$
\{\vec{x}\in\{0,1\}^{n-1}:\ i\in f(\vec{x})\}
$  
into disjoint pairs of the form $\{\vec{x},\vec{x}+e_i\}$. 

Finally, we consider the expression $\sum_{\vec{x}\in \{0,1\}^{n-1}} |f(\vec{x})|$. On one hand, we have that
\begin{equation}
\label{eq:odd}
\begin{split}
\sum_{\vec{x}\in \{0,1\}^{n-1}} |f(\vec{x})| &= \sum_{i\in N} |\{\vec{x}\in\{0,1\}^{n-1}:\ i\in f(\vec{x})\}|\\
& = \left(2^{n-1}-1\right) + \sum_{i\in N\setminus\{n\}} |\{\vec{x}\in\{0,1\}^{n-1}:\ i\in f(\vec{x})\}|,
\end{split}
\end{equation}
where the second equality is obtained by separating $|\{\vec{x}\in\{0,1\}^{n-1}:\ n\in f(\vec{x})\}|$ from the sum, and observing that it follows from the first two constraints that this expression equals $2^{n-1}-1$. Since $2^{n-1}-1$ is odd and $\sum_{i\in N\setminus\{n\}} |\{\vec{x}\in\{0,1\}^{n-1}:\ i\in f(\vec{x})\}|$ is even,~\eqref{eq:odd} implies that $\sum_{\vec{x}\in \{0,1\}^{n-1}} |f(\vec{x})|$ is odd. 

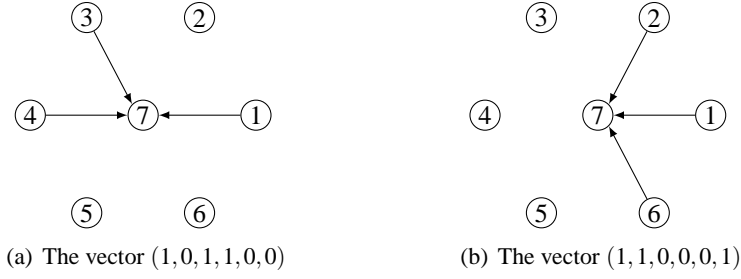
\begin{figure}[t]
\centering
\subfigure[The vector $(1,0,1,1,0,0)$]{%
\begin{tikzpicture}[scale=1.5]

\tikzstyle{blackdot}=[circle,draw=black,fill=black,thin,
inner sep=0pt,minimum size=1.5mm]
\tikzstyle{whitedot}=[circle,draw=black,fill=white,thin,
inner sep=0pt,minimum size=4mm]

\node (7) at (0,0) [whitedot] {\small{7}}; 

\node (1) at (0:1cm) [whitedot] {\small{1}};
\node (2) at (60:1cm) [whitedot] {\small{2}};
\node (3) at (120:1cm) [whitedot] {\small{3}};
\node (4) at (180:1cm) [whitedot] {\small{4}};
\node (5) at (240:1cm) [whitedot] {\small{5}};
\node (6) at (300:1cm) [whitedot] {\small{6}};

\node at (-1.5,0) {};
\node at (1.5,0) {};

\draw [-latex] (1) -- (7.east);
\draw [-latex] (3) -- (7.north west);
\draw [-latex] (4) -- (7.west);

%
%
%

\end{tikzpicture}}
\hspace{1cm}
\subfigure[The vector $(1,1,0,0,0,1)$]{%
\begin{tikzpicture}[scale=1.5]

\tikzstyle{blackdot}=[circle,draw=black,fill=black,thin,
inner sep=0pt,minimum size=1.5mm]
\tikzstyle{whitedot}=[circle,draw=black,fill=white,thin,
inner sep=0pt,minimum size=4mm]

\node (7) at (0,0) [whitedot] {\small{7}}; 

\node (1) at (0:1cm) [whitedot] {\small{1}};
\node (2) at (60:1cm) [whitedot] {\small{2}};
\node (3) at (120:1cm) [whitedot] {\small{3}};
\node (4) at (180:1cm) [whitedot] {\small{4}};
\node (5) at (240:1cm) [whitedot] {\small{5}};
\node (6) at (300:1cm) [whitedot] {\small{6}};

\node at (-1.5,0) {};
\node at (1.5,0) {};

\draw [-latex] (1) -- (7.east);
\draw [-latex] (2) -- (7.north east);
\draw [-latex] (6) -- (7.south east);

%
%
%

\end{tikzpicture}}
\caption{Correspondence between stars and binary $(n-1)$-vectors, for $n=7$} 
\label{fig:star} 
\end{figure}

On the other hand, it trivially holds that 
$$
\sum_{\vec{x}\in \{0,1\}^{n-1}} |f(\vec{x})| = \sum_{\vec{x}\in \{0,1\}^{n-1}} k = 2^{n-1}\cdot k,
$$
hence $\sum_{\vec{x}\in \{0,1\}^{n-1}} |f(\vec{x})|$ is even. We have reached a contradiction.
\end{proof}

It is interesting to note that if we slightly change the problem formulation by allowing the selection of \emph{at most} $k$ agents for $k\geq 2$ then it is possible to design a curious deterministic SP mechanism with a finite approximation ratio that selects at most two agents. The reader is referred to Appendix~\ref{app:leqk} for more details.

\section{Randomized Mechanisms}
\label{sec:rand}

In Section~\ref{sec:det} we have established a total impossibility result with respect to deterministic SP $k$-selection mechanisms. In this section we ask to what extent this result can be circumvented using randomization. 

\subsection{SP Randomized Mechanisms}
\label{subsec:rand_sp}

As we move to the randomized setting, it immediately becomes apparent that Theorem~\ref{thm:det_lb} no longer applies. Indeed, a randomized SP $k$-selection mechanism with a finite approximation ratio can be obtained by simply selecting $k$ agents at random. However, this mechanism still yields a poor approximation ratio. Can we do better?

Consider first a simple deterministic mechanism that partitions the agents into two predetermined subsets $S_1$ and $S_2$. Next, the mechanism discards all edges between pairs of agents in the same subset. Finally, the mechanism chooses the top $k/2$ agents from each subset. In other words, the mechanism selects the $k/2$ agents with highest indegree from each subset, where the indegree is calculated only on the basis of incoming edges from the other subset. This mechanism is clearly SP. Indeed, consider some $i\in S_t$, $t\in \{1,2\}$; its outgoing edges to agents inside its subset are disregarded, whereas its outgoing edges to agents in $S_{3-t}$ can only influence which agents are selected from $S_{3-t}$. However, even without Theorem~\ref{thm:det_lb} it is easy to see that the mechanism does not yield a finite approximation ratio, since it might be the case that the only edges in the graph are between agents in the same subset.

\begin{figure}
\centering
\subfigure[The given graph]{%
\label{sub:a}
\begin{tikzpicture}[scale=1.5]

\tikzstyle{blackdot}=[circle,draw=black,fill=black,thin,
inner sep=0pt,minimum size=1.5mm]
\tikzstyle{whitedot}=[circle,draw=black,fill=white,thin,
inner sep=0pt,minimum size=4mm]

\node (1) at (0:1cm) [whitedot] {\small{1}};
\node (2) at (60:1cm) [whitedot] {\small{2}};
\node (3) at (120:1cm) [whitedot] {\small{3}};
\node (4) at (180:1cm) [whitedot] {\small{4}};
\node (5) at (240:1cm) [whitedot] {\small{5}};
\node (6) at (300:1cm) [whitedot] {\small{6}};

\draw [-latex] (1) -- (2.south east);
\draw [-latex] (3) -- (1.north west);
\draw [-latex] (4) -- (1.west);
\draw [-latex] (4) -- (2.south west);
\draw [-latex] (4) -- (3.south west);
\draw [-latex] (4) -- (5.north west);
\draw [-latex] (4) -- (6.north west);
\draw [-latex] (6) -- (2.south);
\draw [-latex] (6) -- (5.east);

\node at (0,-1.3) {};

\end{tikzpicture}}
\hspace{2cm}
\subfigure[The partitioned graph]{%
\label{sub:b}
\begin{tikzpicture}[scale=1.5]

\tikzstyle{blackdot}=[circle,draw=black,fill=black,thin,
inner sep=0pt,minimum size=1.5mm]
\tikzstyle{whitedot}=[circle,draw=black,fill=white,thin,
inner sep=0pt,minimum size=4mm]

\node (3) at (0,0) [whitedot] {\small{3}};
\node (4) at (0,-1) [whitedot] {\small{4}};
\node (2) at (2,0) [whitedot] {\small{2}};
\node (1) at (2,-1) [whitedot] {\small{1}};
\node (6) at (2,-2) [whitedot] {\small{6}};
\node (5) at (0,-2) [whitedot] {\small{5}};

\draw [-latex] (3) -- (1.north west);
\draw [-latex] (4) -- (1.west);
\draw [-latex] (4) -- (2.west);
\draw [-latex] (4) -- (6.north west);
\draw [-latex] (6) -- (5.east);

\draw[dashed] (0,-1) ellipse (0.5 and 1.4);
\draw[dashed] (2,-1) ellipse (0.5 and 1.4);

\end{tikzpicture}
}
\caption{Example for the Random $2$-Partition Mechanism, with $n=6$ and $k=2$. Figure~\ref{sub:a} illustrates the given graph. The mechanism randomly partitions the agents into two subsets, shown in Figure~\ref{sub:b}, and disregards the edges inside each group. The mechanism then selects the best agent in each group based on the incoming edges from the other group; in the example, the selected subset is $\{1,5\}$, with a sum of indegrees of four, whereas the optimal subset is $\{2,5\}$, with a sum of indegrees of five.}
\label{fig:2group} 
\end{figure}
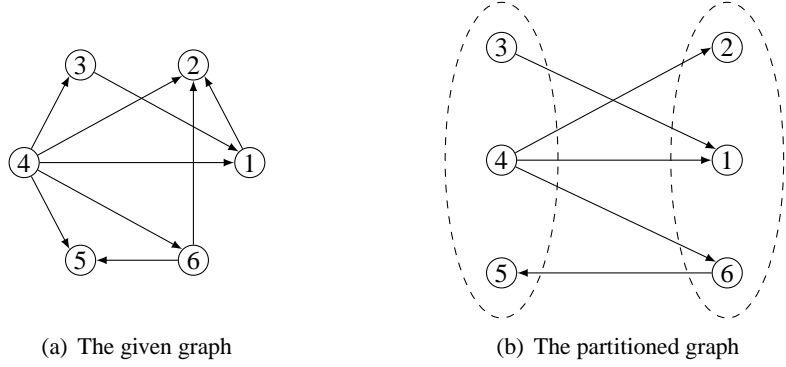

We leverage and refine the partition idea in order to design a randomized SP mechanism that yields a constant approximation ratio. More accurately, we define an infinite family of mechanisms, parameterized by a parameter $m\in \mathbb{N}$. Given $m$, the mechanism randomly partitions the set of agents into $m$ subsets, and then selects (roughly) the top $k/m$ agents from each subset, based only on the incoming edges from agents in other subsets. Below we give a more formal specification of the mechanism; an example can be found in Figure~\ref{fig:2group}. 

\medskip
\noindent\textbf{The Random $m$-Partition Mechanism ($m$-RP)}
\begin{enumerate}
\item Assign each agent independently and uniformly at random to one of $m$ subsets $S_1,\ldots,S_m$.
\item Let $T\subset \{1,\ldots,m\}$ be a random subset of size $k-m\cdot\lfloor k/m\rfloor$.
\item If $t\in T$, select the $\lceil k/m\rceil$ agents from $S_t$ with highest indegrees based only on edges from $N\setminus S_t$. If $t\notin T$, select the $\lfloor k/m\rfloor$ agents from $S_t$ with highest indegrees based only on edges from $N\setminus S_t$. Break ties lexicographically in both cases. If one of the subsets $S_t$ is smaller than the number of agents to be selected from this subset, select the entire subset.
\item If only $k'<k$ agents were selected in Step 3, select $k-k'$ additional agents uniformly from the set of agents that were not previously selected. 
\end{enumerate}

Note that if $k=1$ and $m=2$ then we select one agent from one of the two subsets, based on the incoming edges from the other. In this case, step 2 is equivalent to a toss of a fair coin that determines from which of the two subsets we select an agent.

As in the deterministic case, given a partition of the agents into subsets $S_1,\ldots,S_m$, the choice of agents that are selected from $S_t$ is independent of their outgoing edges. Furthermore, the partition is independent of the input. Therefore, $m$-RP is SP.\footnote{The mechanism is even \emph{universally SP}, see Section~\ref{sec:disc}.} The following theorem explicitly states the approximation guarantees provided by $m$-RP; the technical and rather delicate proof of the theorem is relegated to Appendix~\ref{app:2group}. 

\begin{theorem}
\label{thm:2group}
Let $N=\{1,\ldots,n\}$, $k\in\{1,\ldots,n-1\}$. For every value of $m$, $m$-RP is SP. Furthermore,
\begin{enumerate}
\item $2$-RP has an approximation ratio of four, and
\item $\left(\left\lceil k^{1/3}\right\rceil\right)$-RP has an approximation ratio of $1+\bigO(1/k^{1/3})$.
\end{enumerate} 
\end{theorem}

In fact, we can choose the best value of $m$ for any given value of $k$ when we apply $m$-RP. In other words, Theorem~\ref{thm:2group} implies that for every $k$ there exists an SP mechanism with an approximation ratio of $\min\{4,1+\bigO(1/k^{1/3})\}$, that is, an approximation ratio that is bounded from above by four for any value of $k$, and approaches one as $k$ grows. 

It follows from the theorem that, for $k=1$, $2$-RP has an approximation ratio of four; for this case $m$-RP with $m>2$ has a strictly worse ratio. It is interesting to note that the analysis is tight. Indeed, consider a graph $G=(N,E)$ with only one edge from agent $1$ to agent $n$, that is, $E=\{(1,n)\}$. Assume without loss of generality that agent $n$ is assigned to $S_1$. In order for agent $n$ to be selected, two events must occur:
\begin{enumerate}
\item $T=\{1\}$, that is, the winner must be selected from $S_1$. This happens with probability $1/2$.
\item Either $1\in S_2$, or $|S_1|=1$. The probability that $1\in S_2$ is $1/2$. The probability that $|S_1|=1$, given that $n\in S_1$, is $1/2^{n-1}$. By the union bound, the probability of this event is at most $1/2+1/2^{n-1}$. 
\end{enumerate}
It is clear that $n$ cannot be selected unless the first event occurs. If the second event does not occur, it follows that $n$ has an indegree of zero based on the incoming edges from $S_2$, and there are other alternatives in $S_1$ (which also have an indegree of zero). Since tie-breaking is lexicographic, agent $n$ would not be selected. As the two events are independent, the probability of both occurring is therefore at most $1/4+1/2^n$. We conclude that the approximation ratio of the mechanism cannot be smaller than 
$$
\frac{1}{\left(\frac{1}{4}+\frac{1}{2^n}\right)\cdot 1} = 4 - \bigO\left(\frac{1}{2^n}\right).
$$

We next provide a very simple, though rather weak, lower bound for the approximation ratio yielded by randomized SP $k$-selection mechanisms. Let $k\in\{1,\ldots,n-1\}$, and let $f:\Gcal\rightarrow \Delta(\Scal_k)$ be a randomized SP $k$-selection mechanism. Consider the graph $G=(N,E)$ where 
$$
E = \{(i,i+1):\ i=1,\ldots,k\}\cup \{(k+1,1)\},
$$
i.e., $E$ is a directed cycle on the agents $1,\ldots,k+1$. Then there exists an agent $i\in\{1,\ldots,k+1\}$, without loss of generality agent $1$, that is included in $f(G)$ with probability at most $k/(k+1)$. Now, consider the graph $G'$ where $E'=E\setminus \{(1,2)\}$, that is, agent $1$ removes its outgoing edge to agent $2$. By strategyproofness, agent $1$ is included in $f(G')$ with probability at most $k/(k+1)$. Any subset $S\in \Scal_k$ such that $1\notin S$ has at most $k-1$ incoming edges in $G'$. It follows that the expected number of incoming edges in $f(G')$ is at most
$$
\frac{k}{k+1}\cdot k + \frac{1}{k+1}\cdot (k-1) = \frac{k^2+k-1}{k+1}.
$$
Hence the approximation ratio of $f$ cannot be smaller than 
\begin{equation}
\label{eq:rand_lb}
\frac{k}{\frac{k^2+k-1}{k+1}} = 1+\frac{1}{k^2+k-1}.
\end{equation}
We have therefore proved the following easy result.
\begin{theorem}
\label{thm:rand_lb}
Let $N=\{1,\ldots,n\}$, $n\geq 2$, $k\in \{1,\ldots,n-1\}$. Then there is no randomized SP $k$-selection mechanism with an approximation ratio smaller than $1+\Omega(1/k^2)$.
\end{theorem}

Not surprisingly, the lower bound given by Theorem~\ref{thm:rand_lb} converges to one, albeit more quickly than the upper bound of Theorem~\ref{thm:2group}. As usual, an especially interesting special case is when $k=1$. Equation~\eqref{eq:rand_lb} gives an explicit lower bound of two for this case. On the other hand, Theorem~\ref{thm:2group} gives an upper bound of four. We conjecture that the correct value is two.
\begin{conjecture}
\label{conj:2approx}
There exists a randomized SP $1$-selection mechanism with an approximation ratio of two.
\end{conjecture}

One deceptively promising avenue for proving the conjecture is designing an iterative version of the Random Partition Mechanism. Specifically, we start with an empty subset $S\subset N$, and at each step add to $S$ an agent from $N\setminus S$ that has minimum indegree based on the incoming edges from $S$, breaking ties randomly (so, in the first step we would just add to $S$ a random agent). The last agent that remains outside $S$ is selected. This SP mechanism does remarkably well on some difficult instances, but fails spectacularly on a contrived counterexample. We give a formal specification of this \emph{Sliding Partition Mechanism}, and construct the illuminating counterexample, in Appendix~\ref{app:iterative}.

\subsection{GSP Randomized Mechanisms}
\label{subsec:rand_gsp}

In the beginning of Section~\ref{subsec:rand_sp} we identified a trivial randomized SP $k$-selection mechanism, namely the one that selects a subset of $k$ agents at random. Of course this mechanism is even GSP, since the outcome is completely independent of the reported graph. 

We claim that selecting a random $k$-subset gives an approximation ratio of $n/k$. Indeed, consider an optimal subset $K^*\subseteq N$ with $|K^*|=k$. Each agent $i\in K^*$ is included in the selected subset with probability $k/n$, and hence in expectation contributes a $(k/n)$-fraction of its indegree to the expected total indegree of the selected subset. By linearity of expectation, the expected total indegree of the selected subset is at least a $(k/n)$-fraction of the total indegree of $K^*$.  

Theorem~\ref{thm:2group} implies that we can do much better if we just ask for strategyproofness. If one asks for \emph{group} strategyproofness, on the other hand, just selecting a random subset turns out to be optimal up to a tiny gap.
\begin{theorem}
\label{thm:randgsp_lb}
Let $N=\{1,\ldots,n\}$, $n\geq 2$, and let $k\in\{1,\ldots,n-1\}$. No randomized GSP $k$-selection mechanism can yield an approximation ratio smaller than $(n-1)/k$.
\end{theorem}

\begin{proof}
Let $f:\Gcal\rightarrow \Scal_k$ be a randomized GSP mechanism. Given the empty graph, there are two agents $i,j\in N$ such that each is selected with probability at most $k/(n-1)$. 

Consider the graph $G'$ where $E'=\{(i,j), (j,i)\}$, that is, there are only two edges in $G'$, from $i$ to $j$ and from $j$ to $i$. By group strategyproofness, it must hold for either $i$ or $j$ that $f(G')$ selects this agent with probability not greater than under the empty graph; we may assume without loss of generality that $f(G')$ selects $i$ with probability at most $k/(n-1)$. 

Now consider the graph $G''$ with $E''=\{(j,i)\}$. By strategyproofness, $i$ is selected with equal probability under $f(G')$ and $f(G'')$, that is, with probability at most $k/(n-1)$. Since $i$ is the only agent with an incoming edge in $G''$, the approximation ratio is at least $(n-1)/k$. 
\end{proof}

Note that Theorem~\ref{thm:randgsp_lb} holds even if one is merely interested in coalitions of size at most two.

\section{Discussion}
\label{sec:disc}

In this section we discuss several prominent issues, and list some open problems. 

\medskip
\noindent\textbf{Payments.} 
If payments are allowed and the preferences of the agents are quasi-linear then truthful implementation of the optimal solution is straightforward: simply give one unit of payment to each agent that is not selected. This can be refined by only paying ``pivotal'' agents that are not selected, that is, agents that would have been selected had they lied. However, even under the latter scheme we may have to pay all the non-selected agents (e.g., when the graph is a clique). Moreover, a simple argument shows that there is no truthful payment scheme that does better.  

\medskip
\noindent\textbf{The utility model.} 
We have studied an ``extreme'' utility model, where an agent is only interested in the question of its own selection. The restriction of the preferences of the agents allows us to circumvent impossibility results that hold with respect to more general preferences, e.g., the Gibbard-Satterthwaite Theorem~\cite{Gib73,Sat75} and its generalization to randomized rules~\cite{Gib77}.

It is possible to consider a more sensitive utility function, where an agent receives a utility of one if it is selected, plus a utility of $\beta\geq 0$ for each of its (outgoing) neighbors that is selected. In this model the social welfare (sum of utilities) of a set $S$ of selected agents is $k$ plus $\beta$ times the total indegree of $S$. Hence, if $\beta>0$, a set $S$ maximizes the social welfare if and only if it maximizes the total indegree. In particular, if $\beta>0$ and payments are available, we can use the VCG mechanism~\cite{Vick61,Clar71,Grov73} (see~\cite{Nis07} for an overview) to maximize the total indegree in a truthful way. 

It is easy to verify that any upper bound in the $0$--$1$ model (with total indegree as the target function) also holds in the $\beta$--$1$ model (with social welfare as the target function), hence Theorem~\ref{thm:2group} is true for the latter. Furthermore, if not zero, $\beta$ may still be very small in many settings, like those described in Section~\ref{sec:intro}. In such cases a variation on the random partition mechanism achieves an approximation ratio close to one for the social welfare, even when $k=1$. Finally, note that if $\beta\geq 1$ then simply selecting the optimal solution (and breaking ties lexicographically) is SP.


\medskip
\noindent\textbf{Weights and an application to conference reviews.} A seemingly natural generalization of our model can be obtained by allowing weighted edges. Interestingly, our main positive result, namely Theorem~\ref{thm:2group}, also holds in this more general setting (subject to minor modifications to its formulation and proof). However, closer scrutiny reveals that it is our target function that is often meaningless in the weighted setting. Indeed, the absence of an edge between $i$ and $j$ would in this context imply that $i$ has no information about $j$, whereas an edge with small weight would imply that $i$ dislikes or distrusts $j$. Therefore, maximizing the sum of weights on incoming edges may not be desirable. 

That said, in very specific situations maximizing the sum of weights on incoming edges makes perfect sense; one prominent example is conference reviews. In this context the reviewers assign scores to papers while often submitting a paper of their own, and a subset of papers must be selected. This setting is special since it is usually the case that each paper is reviewed by three reviewers, i.e., each agent has exactly three incoming weighted edges, hence maximizing the sum of scores is the same as maximizing the average score. We conclude that $m$-RP can be employed to build a truthful conference program!

\medskip
\noindent\textbf{Universal strategyproofness vs.\@ strategyproofness in expectation.} 
In the context of randomized mechanisms, two flavors of strategyproofness are usually considered. A mechanism is \emph{universally SP} if for every fixed outcome of the random choices made by the mechanism an agent cannot gain by lying, that is, the mechanism is a distribution over SP mechanisms. A mechanism is \emph{SP in expectation} if an agent cannot increase its expected utility by lying. In this paper we have used the latter definition, which clearly is the weaker of the two. On the one hand, this strengthens the randomized SP lower bound of Theorem~\ref{thm:rand_lb}. On the other hand, notice that the randomized mechanisms of Section~\ref{sec:rand} are in fact universally SP. Indeed, for every fixed partition, selecting agents from one subset based on incoming edges from other subsets is SP. Hence, Theorem~\ref{thm:2group} is even stronger than originally stated. 

\medskip
\noindent\textbf{Open problems.} 
Our most enigmatic open problem is the gap for randomized SP $1$-selection mechanisms: Theorem~\ref{thm:2group} gives an upper bound of four, while Theorem~\ref{thm:rand_lb} gives a lower bound of two. We conjecture that there exists a randomized SP $1$-selection mechanism that gives a $2$-approximation. 

In addition, a potentially interesting variation of our problem can be obtained by changing the target function. One attractive option is to maximize the minimum indegree in the selected subset. Clearly, our total impossibility for deterministic SP mechanisms (Theorem~\ref{thm:det_lb}) carries over to this new target function. However, it is unclear what can be achieved using randomized SP mechanisms.

\section{Acknowledgments}

We thank Moshe Babaioff, Liad Blumrosen, Michal Feldman, Gil Kalai, David Parkes, Yoav Shoham, and Aviv Zohar for valuable discussions. 

\bibliographystyle{plain}
\bibliography{abb,ultimate}

\appendix

\section{Proof of Theorem~\ref{thm:2group}}
\label{app:2group}

For the first part of the theorem, consider an optimal set of $k$ agents (which might not be unique), and denote it by $K^*\subseteq N$. Let $\opt$ be the sum of the indegrees of the agents in $K^*$, that is,
$$
\opt = \sum_{i\in K^*} \deg(i).
$$
We wish to show that the mechanism selects a $k$-subset with an expected number of $\opt/4$ incoming edges. 

\begin{figure}[t]
\centering
\begin{tikzpicture}[scale=1.5]

\tikzstyle{whitedot}=[circle,draw=black,fill=white,thin,
inner sep=0pt,minimum size=4mm]

\node (1) at (0,0) [whitedot] {\small{1}};
\node (2) at (1,0) [whitedot] {\small{2}};
\node (3) at (2,0) [whitedot] {\small{3}};
\node (4) at (3,0) [whitedot] {\small{4}};

\node (5) at (-1,1) [whitedot] {\small{5}};
\node (6) at (-1,-1) [whitedot] {\small{6}};%
\node (7) at (4,1) [whitedot] {\small{7}};
\node (8) at (4,-1) [whitedot] {\small{8}};%

\draw [-latex] (5) -- (1.north west);
\draw [-latex] (5) -- (2.north west);
\draw [-latex] (6) -- (1.south west);
\draw [-latex] (6) -- (2.south west);

\draw [-latex] (7) -- (3.north east);
\draw [-latex] (7) -- (4.north east);
\draw [-latex] (8) -- (3.south east);
\draw [-latex] (8) -- (4.south east);

\draw [-latex] (2) to [bend left] (7.west);
\draw [-latex] (2) to [bend right] (8.west);

\draw [-latex] (3) to [bend right] (5.east);
\draw [-latex] (3) to [bend left] (6.east);

\draw [-latex] (2) to [bend left] (3.west);
\draw [-latex] (3) to [bend left] (2.east);

\draw[dashed] (-1.3,1.3) rectangle (1.4,-1.3);

\draw[dashed] (4.3,1.3) rectangle (1.6,-1.3);

\node at (4,0) {$S_2$};
\node at (-1,0) {$S_1$};

\draw[dashed] (1.5,0) ellipse (1.9 and 0.5);

\node at (0.5,0) {$K_1^*$};
\node at (2.5,0) {$K_2^*$};


\end{tikzpicture}
\caption{An illustration of the proof of Theorem~\ref{thm:2group}, for $n=8$ and $k=4$. In the given graph $G$, the optimal subset is $K^*=\{1,2,3,4\}$. $N$ is partitioned into $S_1=\{1,2,5,6\}$ and $S_2=\{3,4,7,8\}$, which partitions $K^*$ into $K^*_1=\{1,2\}$ and $K^*_2=\{3,4\}$. We have that $d_1=d_2=1$.}
\label{fig:kstar}
\end{figure}
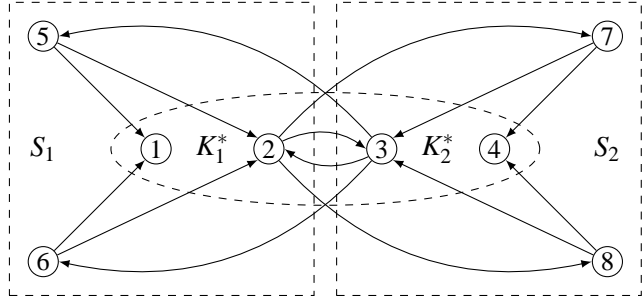

Consider some partition $\pi$ of the agents into two subsets $S_1$ and $S_2$. In particular, let $K^*$ be partitioned into $K_1^*\subseteq S_1$ and $K_2^*\subseteq S_2$, and assume without loss of generality that $|K_1^*|\geq |K_2^*|$. Denote by $d_1$ the number of edges from $S_2$ to $K_1^*$, that is, 
$$
d_1 = |\{(i,j)\in E:\ i\in S_2\wedge j\in K_1^*\}|,
$$ 
and similarly
$$
d_2 = |\{(i,j)\in E:\ i\in S_1\wedge j\in K_2^*\}|.
$$ 
See Figure~\ref{fig:kstar} for an illustration.

Note that step 2 of the $2$-RP mechanism is equivalent to flipping a fair coin to determine whether we select $\lceil k/2\rceil$ agents from $S_1$ and $\lfloor k/2\rfloor$ agents from $S_2$ (when $T=\{1\}$), or vice versa (when $T=\{2\}$). Now, since $|K_2^*|\leq \lfloor k/2\rfloor$ (by our assumption that $|K_1^*|\geq |K_2^*|$), it follows that the subset of $S_2$ selected by the mechanism has at least $d_2$ incoming edges, regardless of whether $T=\{1\}$ or $T=\{2\}$, and even if $|S_2|<\lfloor k/2 \rfloor$. Moreover, since $|K_1^*|\leq |K^*|=k$ it holds that the subset of $S_1$ selected by the mechanism has at least $(\lceil k/2\rceil/k)\cdot d_1$ incoming edges if $T=\{1\}$, and at least $(\lfloor k/2\rfloor/k)\cdot d_1$ if $T=\{2\}$.
Therefore, we have that
\begin{equation}
\label{eq:pointwise}
\begin{split}
\E{\mech\ |\ \pi} & = \E{\mech\ |\ \pi\wedge T=\{1\}}\cdot \frac{1}{2} + \E{\mech\ |\ \pi\wedge T=\{2\}}\cdot \frac{1}{2}\\
& \geq \left(\frac{\lceil k/2\rceil}{k}\cdot d_1 + d_2\right)\cdot \frac{1}{2} + \left(\frac{\lfloor k/2\rfloor}{k}\cdot d_1 + d_2\right)\cdot \frac{1}{2}\\
& = \frac{d_1}{2} + d_2\geq \frac{d_1+d_2}{2}.
\end{split}
\end{equation}

For a random partition of the agents into $S_1$ and $S_2$, each edge has probability $1/2$ of being an edge between the two subsets, and probability $1/2$ of being inside one of the subsets. Hence, by linearity of expectation, the expected number of edges incoming to $K^*$ that are between the two subsets is $\opt/2$. Formally, for a partition $\pi$, let $S_1^{\pi}$ and $S_2^{\pi}$ be the two subsets of agents, and let 
$$
d^{\pi} = |\{(i,j)\in E:\ (i\in S_1^{\pi}\wedge j\in S_2^{\pi}\cap K^*)\vee (i\in S_2^{\pi}\wedge j\in S_1^{\pi}\cap K^*)\}|.
$$ 
Then it holds that 
\begin{equation}
\label{eq:exp}
\sum_{\pi} \Pr{\pi}\cdot d^{\pi} = \frac{\opt}{2}.
\end{equation}
We can now conclude that
$$
\E{\mech} = \sum_{\pi} \E{\mech\ |\ \pi}\cdot\Pr{\pi}
 \geq  \sum_{\pi} \Pr{\pi}\cdot \frac{d^\pi}{2}
 = \frac{\opt}{4},
$$
where the second transition follows from~\eqref{eq:pointwise} and the third transition follows from~\eqref{eq:exp}. 

We now turn to the second part of the theorem. For ease of exposition, we will omit the various floors and ceilings from the proof, as we are looking for an asymptotic result. We employ one additional idea: if $k$ is large enough, the random partition into $k^{1/3}$ subsets will be relatively balanced. A direct approach would be to bound the probability that the number of optimal agents in some subset deviates significantly from $k^{2/3}$, and then proceed in a way similar to the first part. We however take a somewhat different approach that yields a better result.

Consider the agents in the optimal set $K^*$, and assume without loss of generality that $K^*=\{1,\ldots,k\}$. Given $i\in K^*$, we define a random variable $Z_i$ that depends on the random partition of $N$ to $S_1,\ldots,S_{k^{1/3}}$ as follows:
$$
Z_i = |\{j\in K^*\setminus\{i\}:\ \exists t\ \text{s.t.}\ i\in S_t\wedge j\in S_t\}|,
$$
that is, $Z_i$ is the number of agents in the optimal set, excluding $i$ itself, that are in the same random subset as agent $i$. We have
\begin{equation}
\label{eq:calc1}
\E{\mech} = \sum_{s_1,\ldots,s_k} \E{\mech\ |\ Z_1=s_1,\ldots,Z_k=s_k}\cdot \Pr{Z_1=s_1,\ldots,Z_k=s_k},
\end{equation}
where the probability is taken over random partitions. 

Recall that the $k^{1/3}$-RP Mechanism selects the top $k^{2/3}$ agents from each subset. Let 
$$
\sigma_s = \min\{1,k^{2/3}/(s+1)\}.
$$ 
Furthermore, given $i\in K^*$ and a partition, let  
$$
d'_i = |\{(j,i)\in E:\ j\in S_{t_1}\wedge i\in S_{t_2}\wedge t_1\neq t_2 \}|,
$$
i.e., $d'_i$ is the number of edges incoming to agent $i$ from other subsets. Using similar arguments to those employed to obtain~\eqref{eq:pointwise}, we get
\begin{equation}
\label{eq:calc2}
\begin{split}
\E{\mech\ |\ Z_1=s_1,\ldots,Z_k=s_k}&\geq \E{\sum_{i\in K^*} d'_i\sigma_{s_i}\ |\ Z_1=s_1,\ldots,Z_k=s_k}\\
& =  \sum_{i\in K^*} \E{d'_i\sigma_{s_i}\ |\ Z_1=s_1,\ldots,Z_k=s_k}.
\end{split}
\end{equation}

We wish to obtain an explicit expression for $\E{d'_i\sigma_{s_i}\ |\ Z_1=s_1,\ldots,Z_k=s_k}$.  For $i\in N$ and $S\subseteq N$, let
$$
\deg(i,S) = |\{(j,i)\in E:\ j\in S \}|
$$
be the indegree of agent $i$ based on incoming edges from agents in $S$. We claim that
\begin{equation}
\label{eq:calc3}
\E{d'_i\sigma_{s_i}\ |\ Z_1=s_1,\ldots,Z_k=s_k} = \left( \frac{k-1-s_i}{k-1} \cdot \deg(i,K^*) + \frac{k^{1/3}-1}{k^{1/3}}\cdot\deg(i,N\setminus K^*)\right)\cdot \sigma_{s_i}.
\end{equation}
Indeed, this identity is obtained by using linearity of expectation twice, as any fixed agent in $K^*$ is not in the same subset as agent $i$ with probability $(k-1-s_i)/(k-1)$, and any fixed agent in $N\setminus K^*$ is not in the same subset as agent $i$ with probability $(k^{1/3}-1)/k^{1/3}$. Notice that the expression on the right hand side of~\eqref{eq:calc3} is independent of $s_j$ for all $j\neq i$.

Combining~\eqref{eq:calc1},~\eqref{eq:calc2}, and~\eqref{eq:calc3}, and reversing the order of summation, we conclude that
\begin{equation*}
\begin{split}
\E{\mech} &\geq \sum_{i\in K^*}\sum_{s_1,\ldots,s_k} \Pr{Z_1=s_1,\ldots,Z_k=s_k} \cdot \\ & \hspace{2cm} \left( \frac{k-1-s_i}{k-1} \cdot \deg(i,K^*) + \frac{k^{1/3}-1}{k^{1/3}}\cdot\deg(i,N\setminus K^*)\right)\cdot \sigma_{s_i} \\
& = \sum_{i\in K^*}\sum_{s=0}^{k-1}\Pr{Z_i=s}\cdot \left( \frac{k-1-s}{k-1} \cdot \deg(i,K^*) + \frac{k^{1/3}-1}{k^{1/3}}\cdot\deg(i,N\setminus K^*)\right)\cdot \sigma_s\\
& = \sum_{i\in K^*}\sum_{s=0}^{k-1}\Pr{Z_i=s}\cdot \frac{k-1-s}{k-1}\cdot\deg(i,K^*) \cdot \sigma_s + \\ & \hspace{2cm} \sum_{i\in K^*}\sum_{s=0}^{k-1}\Pr{Z_i=s}\cdot \frac{k^{1/3}-1}{k^{1/3}}\cdot\deg(i,N\setminus K^*)\cdot \sigma_s.
\end{split}
\end{equation*}
On the other hand, we have that
$$
\opt = \sum_{i\in K^*} \left(\deg(i,K^*) + \deg(i,N\setminus K^*)\right) = \sum_{i\in K^*}\deg(i,K^*) + \sum_{i\in K^*}\deg(i,N\setminus K^*).
$$
In order to complete the proof it therefore suffices to prove that for every $i\in K^*$,
\begin{equation}
\label{eq:suff1}
\sum_{s=0}^{k-1}\Pr{Z_i=s}\cdot\frac{k^{1/3}-1}{k^{1/3}}\cdot \sigma_s = 1-\bigO\left(\frac{1}{k^{1/3}}\right),
\end{equation}
and
\begin{equation}
\label{eq:suff2}
\sum_{s=0}^{k-1}\Pr{Z_i=s}\cdot\frac{k-1-s}{k-1}\cdot \sigma_s = 1-\bigO\left(\frac{1}{k^{1/3}}\right).
\end{equation}
Using these equalities we may conclude that
\begin{align*}
\frac{\opt}{\E{\mech}} &\leq \frac{\sum_{i\in K^*}\deg(i,K^*) + \sum_{i\in K^*}\deg(i,N\setminus K^*)}{\sum_{i\in K^*}\left(1-\bigO\left(\frac{1}{k^{1/3}}\right)\right)\deg(i,K^*) + \sum_{i\in K^*}\left(1-\bigO\left(\frac{1}{k^{1/3}}\right)\right)\deg(i,N\setminus K^*)}\\
&= \frac{1}{\left(1-\bigO\left(\frac{1}{k^{1/3}}\right)\right)}
= 1+\bigO\left(\frac{1}{k^{1/3}}\right).
\end{align*}

Since $\sigma_s=1$ for all $s\leq k^{2/3}-1$, in order to establish~\eqref{eq:suff1} we must show that
$$
\sum_{s=k^{2/3}}^{k-1}\Pr{Z_i=s}\cdot\frac{s+1-k^{2/3}}{s+1} = \bigO\left(\frac{1}{k^{1/3}}\right).
$$
Indeed, 
\begin{equation}
\label{eq:series}
\begin{split}
\sum_{s=k^{2/3}}^{k-1}\Pr{Z_i=s}\cdot\frac{s+1-k^{2/3}}{s+1}
&\leq \sum_{x=1}^{2\sqrt{\log k}} \Pr{Z_i\geq k^{2/3} + (x-1)k^{1/3}}\cdot\frac{xk^{1/3}+1}{k^{2/3}+xk^{1/3}+1}\\
& + \Pr{Z_i\geq k^{2/3} + 2\sqrt{\log k}\cdot k^{1/3}}\cdot 1.
\end{split}
\end{equation}

In order to bound the probabilities on the right hand side of~\eqref{eq:series} we employ the following version of the Chernoff bounds~(see, e.g.,~\cite{AS00}, Theorem A.1.11).
\begin{lemma}
\label{thm:cher}
Let $X_1, ..., X_k$ be i.i.d.\@ Bernoulli trials, $\Pr{X_i=1}=p$ for $i=1,\ldots,k$, and denote $X=\sum_{i=1}^k X_i$. In addition, let $\lambda>0$. Then
$$
\Pr{X-kp\geq \lambda}\leq \exp\left( -\frac{\lambda^2}{2kp} + \frac{\lambda^3}{2(kp)^2} \right).
$$
\end{lemma}
$Z_i$ is in fact the sum of $k-1$ i.i.d.\@ Bernoulli trials, but we can safely assume that it is the sum of $k$ trials if we are interested in an upper bound on the probability of the sum being greater than some given value. Using Lemma~\ref{thm:cher} with $\lambda=xk^{1/3}$ and $p=1/k^{1/3}$ we get
\begin{equation}
\label{eq:prob}
\Pr{Z_i\geq k^{2/3} + (x-1)k^{1/3}}\leq  \exp\left( -\frac{(x-1)^2k^{2/3}}{2k^{2/3}} + \frac{(x-1)^3k}{2k^{4/3}} \right) \leq \exp\left(-\frac{(x-1)^2}{4}\right),
\end{equation}
where the second inequality holds for a large enough $k$. Similarly,
$$
\Pr{Z_i\geq k^{2/3} + 2\sqrt{\log k}\cdot k^{1/3}}\leq  \exp\left( -\frac{4k^{2/3}\log k}{2k^{2/3}} + \frac{8k(\log k)^{3/2}}{2k^{4/3}} \right) \leq \exp(-\log k)\leq \frac{1}{k}.
$$
We conclude that the expression on the right hand side of~\eqref{eq:series} is bounded from above by
\begin{align*}
\sum_{x=1}^{2\sqrt{\log k}}\left(\exp\left(-\frac{(x-1)^2}{4}\right)\cdot \frac{xk^{1/3}+1}{k^{2/3}+xk^{1/3}+1}\right) + \frac{1}{k} &\leq \frac{1}{k^{1/3}}\sum_{x=1}^{2\sqrt{\log k}}\left(\exp\left(-\frac{(x-1)^2}{4}\right)\cdot 2x\right) + \frac{1}{k}\\
& = \bigO\left(\frac{1}{k^{1/3}}\right),
\end{align*}
which follows from the fact that the series $\sum_{x=1}^{\infty} \exp(-\Theta(x^2))\cdot \Theta(x)$ converges. This establishes~\eqref{eq:suff1}. 

The proof of~\eqref{eq:suff2} is similar to that of~\eqref{eq:suff1}. It is sufficient to show that
\begin{align*}
\sum_{s=0}^{k^{2/3}-1} \Pr{Z_i=s} \cdot \frac{s}{k-1} &+ \sum_{s=k^{2/3}}^{k^{2/3}+2\sqrt{\log k}\cdot k^{1/3}-1} \Pr{Z_i=s}\left(1-\frac{k-1-s}{k-1}\cdot\frac{k^{2/3}}{s+1}\right)\\
& +\Pr{Z_i\geq k^{2/3}+2\sqrt{\log k}\cdot k^{1/3}}\cdot 1 = \bigO\left(\frac{1}{k^{1/3}}\right).
\end{align*}
It holds that 
$$
\sum_{s=0}^{k^{2/3}-1} \Pr{Z_i=s} \cdot \frac{s}{k-1}\leq \sum_{s=0}^{k^{2/3}-1} \Pr{Z_i=s} \cdot \frac{k^{2/3}-1}{k-1} =  \bigO\left(\frac{1}{k^{1/3}}\right),
$$
and as before,
$$
\Pr{Z_i\geq k^{2/3}+2\sqrt{\log k}\cdot k^{1/3}}\cdot 1 \leq \frac{1}{k}.
$$ 
Finally,
\begin{align*}
&\sum_{s=k^{2/3}}^{k^{2/3}+2\sqrt{\log k}\cdot k^{1/3}-1} \Pr{Z_i=s}\left(1-\frac{k-1-s}{k-1}\cdot\frac{k^{2/3}}{s+1}\right)\\
& = \sum_{s=k^{2/3}}^{k^{2/3}+2\sqrt{\log k}\cdot k^{1/3}-1} \Pr{Z_i=s}\left(1-\left(1-\bigO\left(\frac{1}{k^{1/3}} \right)\right)\cdot\frac{k^{2/3}}{s+1}\right).
\end{align*}
We can thus bound this sum from above as before using~\eqref{eq:prob}. This completes the proof of Theorem~\ref{thm:2group}.

\section{The Edge Scan Mechanism}
\label{app:leqk}

In Theorem~\ref{thm:det_lb} we have seen that a deterministic SP $k$-selection mechanism cannot give a bounded approximation ratio. We now show that if we are allowed to choose \emph{at most} $k$ agents, for $k\geq 2$, it is possible to design an SP mechanism with a bounded approximation ratio. As noted in Section~\ref{sec:det}, it is sufficient to select a subset with an incoming edge, if one exists. 

\begin{figure}
\begin{center}
\begin{tikzpicture}[scale=1.5]

\tikzstyle{blackdot}=[circle,draw=black,fill=black,thin,
inner sep=0pt,minimum size=1.5mm]
\tikzstyle{whitedot}=[circle,draw=black,fill=white,thin,
inner sep=0pt,minimum size=4mm]

\node (1) at (0,0) [whitedot] {\small{1}};
\node (2) at (1,0) [whitedot] {\small{2}};
\node (3) at (2,0) [whitedot] {\small{3}};
\node (4) at (3,0) [whitedot] {\small{4}};
\node (5) at (4,0) [whitedot] {\small{5}};
\node (6) at (5,0) [whitedot] {\small{6}};

\draw [-latex] (4.north east) to [bend left] (5.north west);
\draw [-latex] (2.north east) to [bend left] (4.north west);
\draw [-latex] (3.south west) to [bend left] (1.south east);
\draw [-latex] (3.north east) to [bend left] (6.north west);
\draw [-latex] (4.south west) to [bend left] (3.south east);

%
%
%

\end{tikzpicture}
\end{center}
\caption{Example for the Edge Scan Mechanism. Given this graph, the mechanism would select agent $4$ in the scan from left to right, and agent $3$ in the scan from right to left, so the subset of agents selected by the mechanism is $\{3,4\}$.} 
\label{fig:lr} 
\end{figure}
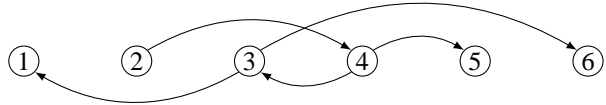

Intuitively, the mechanism, which we refer to as the \emph{Edge Scan Mechanism}, first orders the agents from left to right according to their lexicographic ordering. The mechanism then scans the agents from left to right, until it finds an outgoing edge directed to the right, and selects the agent the edge is pointing at. Similarly, the mechanism scans the agents from right to left until it finds an edge that is directed to the left, and also selects the agent that this edge is pointing at. An example is shown in Figure~\ref{fig:lr}. What follows is a more formal specification of the mechanism.

\medskip
\noindent\textbf{The Edge Scan Mechanism.} 
\begin{enumerate}
\item Partition $E$ into $E_1=\{(i,j)\in E:\ i<j\}$ and $E_2=\{(i,j)\in E:\ i>j\}$.
\item If $E_1\neq \emptyset$, let $i\in N$ be the minimum index such that there exists $j\in N$ with $(i,j)\in E_1$; add to the subset the minimum $j$ such that $(i,j)\in E_1$. Otherwise, add agent $n$ to the subset. 
\item If $E_2\neq \emptyset$, let $i\in N$ be the maximum index such that there exists $j\in N$ with $(i,j)\in E_2$; add to the subset the maximum $j$ such that $(i,j)\in E_2$. Otherwise, add agent $1$ to the subset. 
\end{enumerate}

The Edge Scan Mechanism is clearly SP. Indeed, agent $i$ cannot benefit from adding outgoing edges, since these edges would only point at some other agent. It also cannot benefit from removing outgoing edges. Informally, if the mechanism reaches the point in the scan (from left to right or right to left) where the agent's vote is taken into account, then it is too late for agent $i$ itself to be elected. 

Moreover, either $E_1$ or $E_2$ will contain an edge given that there is at least one edge in the graph, and the Edge Scan Mechanism is guaranteed to select an agent with an incoming edge in this case. It therefore achieves a finite approximation ratio, although this ratio can be as bad as $\Omega(nk)$.

Crucially, the agents selected in both steps of the mechanism can be one and the same; in this case the mechanism would return a singleton subset. A curious implication of Theorem~\ref{thm:det_lb} is that such a selection cannot be completed deterministically and in a strategyproof way to obtain a subset of size two. 

\section{The Sliding Partition Mechanism}
\label{app:iterative}

In this appendix we discuss the \emph{Sliding Partition Mechanism}, informally presented in Section~\ref{sec:rand}. The mechanism is randomized, and was designed to yield an SP upper bound better than four for the $k=1$ case. Although the mechanism ultimately fails in achieving this goal, we believe that the counterexample is surprising and may prove helpful in future attempts to resolve Conjecture~\ref{conj:2approx}. We start with an informal specification of the mechanism.

\medskip
\noindent\textbf{The Sliding Partition Mechanism.}
	\begin{enumerate}
		\item Let $S=\emptyset$.
		\item While $|S|<n-1$, choose $i\notin S$ that has minimum indegree based on edges from agents in $S$, breaking ties randomly. Let $S=S\cup\{i\}$.
		\item Select the agent in $N\setminus \{S\}$.
	\end{enumerate}

When an agent is added to $S$, we say that it is \emph{eliminated}. It is easy to see that this mechanism is SP. Indeed, only the outgoing edges of eliminated agents are taken into account at any stage. Once an agent is eliminated, it no longer has a chance to be selected, therefore it is indifferent to the outcome of the mechanism.

Another interesting observation is that the Sliding Partition Mechanism gives a $2$-approximation for the example where the analysis of the $2$-RP mechanism is tight: a graph with only one edge. Indeed, if $G$ has one edge $(i,j)$, then $j$ is certainly elected once $i$ is eliminated (since then it is the only agent in $N\setminus S$ with an incoming edge from $S$), and $i$ is eliminated before $j$ with probability $1/2$. 

Unfortunately, it is possible to construct a graph where the mechanism does very poorly. For this, consider a tree with agent $1$ at the root. There is a set $T\subset N$ of size $n^{3/5}$ of agents with outgoing edges to $1$, that is, $\deg(1)=n^{3/5}$. In addition, each agent in $T$ has $n^{2/5}$ incoming edges from agents in $N\setminus (\{1\}\cup T)$. The agents in $N\setminus (\{1\}\cup T)$ have an indegree of zero.

Notice that while there are agents in $N\setminus S$ that have no incoming edges from $S$, the mechanism selects one of these agents uniformly at random and eliminates it. Consider the first point in time $t_0$ when all the agents in $T\setminus S$ that were not yet eliminated have at least one incoming edge from $S$; we can assume without loss of generality that at this point agent $1$ has not been eliminated. We claim that if less than $n^{2/5}$ agents from $T$ have been eliminated at time $t_0$, then agent $1$ is guaranteed to be eliminated later on. Indeed, starting at $t_0$, the remaining agents in $N\setminus (\{1\}\cup T)$ are eliminated one after the other (in some random order), because all of them have indegree zero from $S$, while all other remaining agents have at least one incoming edge from $S$. After all the agents in $N\setminus (\{1\}\cup T)$ have been eliminated, each agent in $T\setminus S$ has $n^{2/5}$ incoming edges from $S$. By assumption agent $1$ has less and is eliminated next.

We now claim that with high probability, agent $1$ has less than $n^{2/5}$ incoming edges from $S$ at time $t_0$. Each agent $i\in T$ contributes an edge to $1$ at time $t_0$ if and only if it is eliminated before any of the agents in its incoming neighborhood; this happens with probability roughly $1/n^{2/5}$. Therefore, by linearity of expectation, the expected number of edges from $S$ to agent $1$ at time $t_0$ is roughly $n^{1/5}$. The claim now follows directly from Chernoff's inequality. 

We conclude that the approximation ratio provided by the Sliding Partition Mechanism cannot be smaller than $\Omega(n^{1/5})$. By optimizing the parameters of the example, it is possible to obtain an even stronger lower bound. 

\end{document}